\documentclass[submission,copyright,creativecommons]{eptcs}
\providecommand{\event}{Wivace 2013} 
\usepackage{breakurl}             

\usepackage{amsfonts}
\usepackage{amssymb}
\usepackage{graphicx}
\usepackage{amsmath}
\usepackage{rotating}
\usepackage{tabularx}
\usepackage{setspace} 
\usepackage{color}

\title{Chemical communication between synthetic and natural cells: a possible experimental design.}

\author{Giordano Rampioni \institute{Science Dept. University of Roma Tre, Italy} \email{giordano.rampioni@uniroma3.it}
\and 	Luisa Damiano	\institute{University of Bergamo} \email{luisa.damiano@gmail.com}
\and Marco Messina \institute{Science Dept. University of Roma Tre, Italy} \email{marco.messina@uniroma3.it} 
\and Francesca D'Angelo	\institute{Science Dept. University of Roma Tre, Italy} \email{francesca.dangelo101089@gmail.com}	
\and Livia Leoni \institute{Science Dept. University of Roma Tre, Italy}	\email{livia.leoni@uniroma3.it}
\and Pasquale Stano \institute{Science Dept. University of Roma Tre, Italy}
\email{*Corresponding Author = stano@uniroma3.it	}
}
\def\authorrunning{Rampioni et al.}
\def\titlerunning{Chemical communication between synthetic and natural cells: a possible experimental design}
\begin{document}
\maketitle

\begin{abstract}
The bottom-up construction of synthetic cells is one of the most intriguing and interesting research arenas in synthetic biology. Synthetic cells are built by encapsulating biomolecules inside lipid vesicles (liposomes), allowing the synthesis of one or more functional proteins. Thanks to the \textit{in situ} synthesized proteins, synthetic cells become able to perform several biomolecular functions, which can be exploited for a large variety of applications. This paves the way to several advanced uses of synthetic cells in basic science and biotechnology, thanks to their versatility, modularity, biocompatibility, and programmability. 
In the previous WIVACE (2012) we presented the state-of-the-art of semi-synthetic minimal cell (SSMC) technology and introduced, for the first time, the idea of chemical communication between synthetic cells and natural cells. The development of a proper synthetic communication protocol should be seen as a tool for the nascent field of bio/chemical-based Information and Communication Technologies (bio-chem-ICTs) and ultimately aimed at building soft-wet-micro-robots. 
In this contribution (WIVACE, 2013) we present a blueprint for realizing this project, and show some preliminary experimental results. We firstly discuss how our research goal Ð based on the natural capabilities of biological systems to manipulate chemical signals Ð finds a proper place in the current scientific and technological contexts. Then, we shortly comment on the experimental approaches from the viewpoints of \textit{(i)} synthetic cell construction, and \textit{(ii)} bioengineering of microorganisms, providing up-to-date results from our laboratory. Finally, we shortly discuss how autopoiesis can be used as a theoretical framework for defining synthetic minimal life and minimal cognition, as well as a bridge between synthetic biology and artificial intelligence. 
\end{abstract}

\section{Background}
The goal of synthetic biology (SB), a recently emerging biology branch derived from the combination of biology and engineering, is the programmable construction of biological parts, devices and systems to perform useful functions. In addition to top-down approaches, based on the repurposing of existing cell by genetic and metabolic engineering \cite{5}, a radically different bottom-up approach foresees the assembly of artificial cells from separated molecular parts. Among the various possible designs, the so-called ``semi-synthetic{"} approach \cite{12} appears to be the most promising in terms of feasibility, versatility, modularity, robustness, and possibility of interfacing with biological systems. Semi-synthetic minimal cells (SSMCs) are defined as those synthetic cell-like systems based on the encapsulation of the minimal number of biomolecules (such as nucleic acids, proteins, etc.) inside lipid vesicles (liposomes) capable of displaying minimal living-like properties, like self-maintenance or self-reproduction (Figure 1a). The ultimate goal of SSMC research is the construction of a living synthetic cell. Although this final goal seems Ð to date Ð quite difficult to reach, SSMC technology is rapidly progressing, and a great interest grew around it, from the view-points of origin-of-life, synthetic biology, artificial life, artificial intelligence, systems chemistry, self-organization, and complexity theories.
We have recently proposed that SSMCs can play a major role in the nascent field of bio-chemical Information and Communication Technologies (ICTs) \cite{24}. In particular, inspired by the natural signal processing ability of biological cells, we foresee that chemical information could be manipulated by SSMCs in a programmable way, by reconstructing the minimal set of molecular sensor, actuators, controllers inside liposomes. In this way, by providing SSMCs with the devices required for processing chemical signals, we aim at: \textit{(i)} further advancing the SSMC technology, \textit{(ii)} approaching from the experimental viewpoint the issue of ``minimal cognition{"}, \textit{(iii)} creating a biotechnological tool for advanced drug delivery (and others) applications based on bio-chemical-based ICTs (shortly: bio-chem-ICTs).
Recent theoretical analyses and experimental progress concurred to define our approach, that aims at creating synthetic cells capable of communicating with natural cells and with other synthetic cells via biochemical signals (Figure 1b). The relevance of inter- and intra-cellular molecular communication as examples of bio-chem-ICTs has been put forward by the Suda-Nakano groups \cite{16}, whereas a first example of synthetic communication based on liposomes was published by Davis and collaborators \cite{6}. The basic implication for artificial life of synthetic/natural communication protocols was instead emphasized in a perspective paper \cite{2}, where a hypothetical synthetic cell-based Turing-like test is discussed. Finally, we were further inspired by LeDuc's nanofactories, which are programmable synthetic cells for future nanomedicine applications \cite{10}.
In our previous WIVACE contribution \cite{25} we introduced the main idea of synthetic cell-natural cell chemical communication. We described the state-of-the-art of SSMCs technology and how basic molecular circuitries could be ``plugged-in{"} in the SSMC chassis, which is currently based on protein synthesis. Here we discuss with more details the rationale design underlying the experimental phase, a sort of blueprint commented and supported by preliminary data collected in our laboratories. We also provide a short remark on theoretical foundations of minimal cognitive systems from an autopoietic perspective.

\section{Chemical communication between synthetic and natural cells}
Properly designed man-made molecular systems could communicate with living biological systems, and viceversa, \textit{via} the exchange of chemical species (messengers). In order to define and establish a communication channel, the partners should have, respectively, an encoding/sender system and a decoding/receiver one; moreover, it should be specified how the signal is transferred from the sender to the receiver. In contrast to classical ICTs, based on electrical signals, bio-chem-ICTs make use of chemicals, and therefore the encoding/sending/transmitting/decoding/receiving operations must operate in the molecular domain.
Signal molecules are typically synthesized by enzymes, and this process is often regulated by tuning the amount/activity of enzymes available in the cell. Some signal molecules require complex multi-steps reaction sequences. The signal molecule is often secreted by living cells trough trans-membrane protein machines. However, in some cases the signaling molecule is able to freely cross the biological membrane and move from the sender to the receiver by simple diffusion in aqueous solution.
The last two steps are reception and decodification of the signal. In many cases the receptor protein (which has high affinity for the cognate signal molecule) is located across the biological membrane, the binding of the signal molecule to the receptor causes a conformational change in the protein that transduces the signal to a cytoplasmic effector located inside the cell, which in turn starts an action (response). In this respect, the mechanisms can have different complexity because signal transduction is often a step-wise process mediated by multiple proteins. Actually, some receptors are bifunctional proteins that, upon signal molecule perception, directly activate a response and trigger the desired effect.
It results that in order to design a minimal communication mechanism inspired to cell communication ``protocols{"}, a proper design in terms of choice of molecular parts and devices to be implemented in the synthetic cell is needed. In turn, this is linked to the specific way in which molecular communications should occur. In fact, three kind of practical implementation of chemical communication involving synthetic and natural cells can be envisaged: \textit{(i)} synthetic-to-natural, \textit{(ii)} natural-to-synthetic, and \textit{(iii)} synthetic-to-synthetic. Clearly, different technical and theoretical consequences stem from these three cases and from their combinations.
As we have anticipated, the ``synthetic partner{"} and the ``natural partner{"} of the communication should be endowed with specific functions, and in the next sections we will argue on how these functions can be implemented in synthetic and natural cells, providing a sort of blueprint for a synthetic minimal communication protocol.

\begin{figure}[htbp]
\begin{center}
\includegraphics[width=10cm]{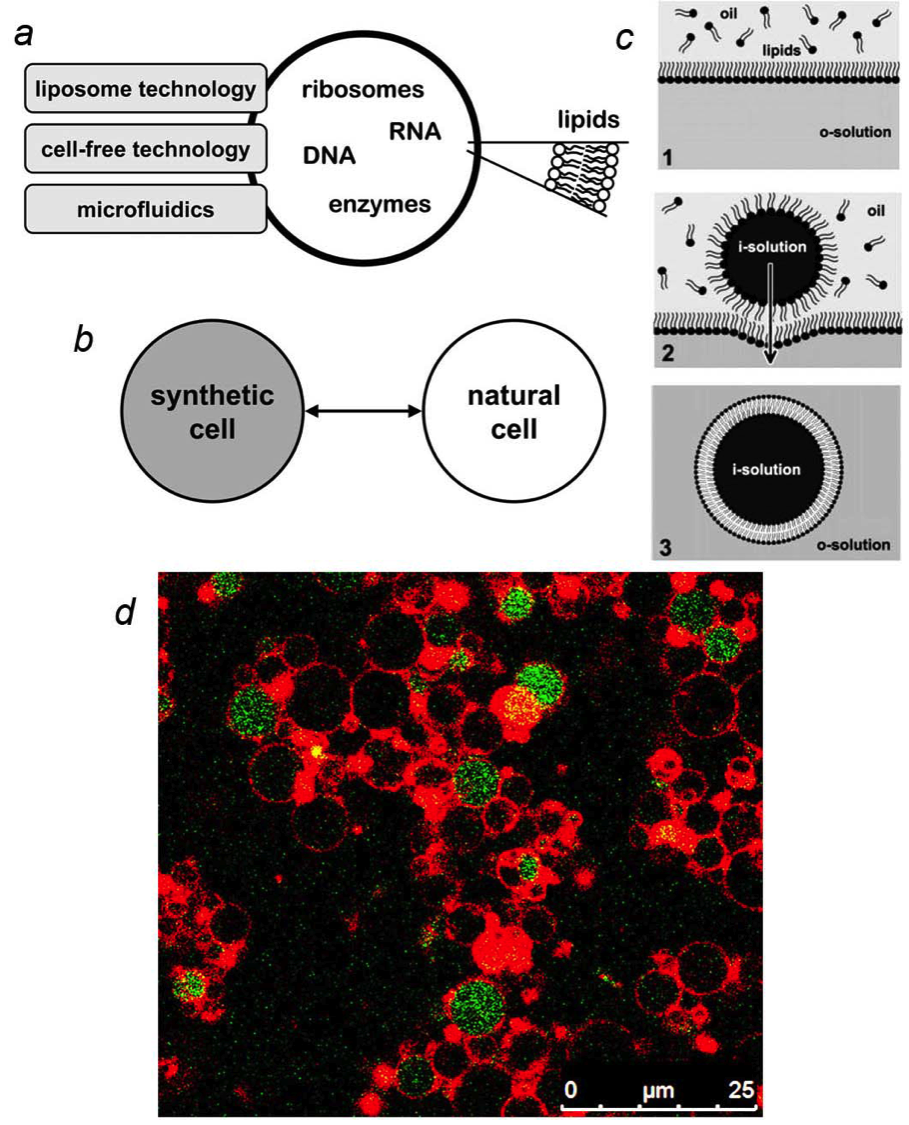}

\caption{(a) Semi-synthetic minimal cells, as derived from the combination of liposome technology, cell-free technology, and possibly microfluidics. (b) Synthetic cells can communicate with natural cells, via exchanging chemical signals. (c) The ``droplet transfer{"} method allows the construction of giant vesicles (GVs) with high encapsulation efficiency (reproduced with minor modifications from \cite{18}, with the permission of Cell Press Elsevier). (d) GVs prepared by the droplet transfer method, visualized by confocal microscopy. The lipid membranes have been stained by Nile Red, whereas the internal green fluorescence is due to the \textit{in situ} produced green fluorescent protein, thanks to the encapsulation of the PURE system and of a plasmid carrying the \textit{gfp} gene.}
\label{figure1}
\end{center}
\end{figure}

\section{Rational design and preliminary results}
As it will be detailed below, we are currently developing SSMCs able to communicate with living cells, and vivecersa, using bacterial communication as biological model system. This choice was due to the fact that the most plausible signal molecule candidates are \textit{N}-acyl-homoserine lactones (AHLs), a well-known class of compounds involved in a widespread bacterial intercellular communication system known as quorum sensing. The possibilities and the constraints related to this choice will be discussed below, by referring to the synthetic partner (SSMCs), and to the natural partner (bacteria).

\subsection{Semi-synthetic minimal cells}
The technology of SSMCs construction derives from the convergence of cell-free technology and liposome technology (more recently, also microfluidic technology impacted the field, by
providing the first devices apt to prepare liposomes in a controllable and reproducible way). In practice, cell-free systems of variable complexity are introduced inside liposomes, with the additional possibility of functionalizing the lipid membrane, for example with membrane proteins or other artificial moieties. Liposomes are cell-like compartments formed by lipid self-assembly. According to the preparation method, it is possible to prepare ``conventional{"} liposomes, whose diameter varies around 100-400 nm, or ``giant{"} liposomes (know as giant vesicles GVs), which can be directly visualized by microscopy (typical size: 2-20 $\mu$m). From the viewpoint of the preparation methods, we can instead distinguish between ``spontaneous{"} and ``directed{"} methods. In the first case, the liposomes form spontaneously, and the process of solute encapsulation (the key event for the construction of a SSMC) also follows a spontaneous route. Despite the quite intriguing recent report on the spontaneous formation of ``super-filled{"} conventional vesicles \cite{11}, the methods that rely on spontaneous self-assembly rarely bring about a homogeneous and reproducible liposome formation (independently from their size), and it is therefore useful to follow ``directed{"} procedures. GVs can be effectively prepared by the droplet-transfer method \cite{17}. In this method, the solutes of interest are firstly emulsified in oil, forming microscopic water droplets surrounded by lipids. Next, the droplets are transformed into vesicles by let them cross an oil/water interface, where the second lipid layer is assembled (Figure 1c). By this procedure, it is possible to construct synthetic cells with very high entrapment efficiency, so that even complex multimolecular cell-free machineries are incorporated inside liposomes. In our hand, the method produces thousands of GVs with encapsulation efficiency around 40\%, as determined by a study on the yields of capture and release of fluorescent probes. Interestingly, we recently demonstrated that GVs can be prepared also in the presence of media typically used to grow bacteria, like Luria Bertani broth (LB) \cite{19}, although a certain degree of aggregation was observed. Overall, the droplet transfer method appears to be the most valuable protocol for liposome preparation when complex multimolecular systems are used, as those required for implementing synthetic communication.
However, due to the large size of GVs, a further size reduction step might be necessary when vesicle size matters. For example, liposomes for intelligent drug delivery applications (\textit{i.e.}, LeDuc's nanofactories) need to be small if vehiculated by bloodstream. A possible route to the preparation of solute-filled small liposomes, that avoids the ``spontaneous{"} hydration processes, could start from GVs prepared by the droplet transfer method and successively sized down by extrusion.

\begin{figure}[htbp]
\begin{center}
\includegraphics[width=10cm]{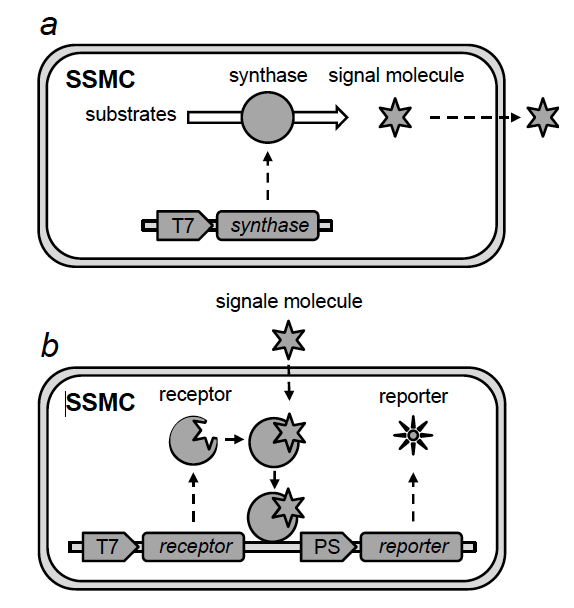}

\caption{Schematic representations of (a) a SSMCs able to synthesize a signal molecule, and of (b) a SSMC able to perceive and respond to a signal molecule. T7, constitutive promoter activated by the T7 RNA polymerase; PS, promoter activated by the signal molecule/receptor complex.}
\label{figure2}
\end{center}
\end{figure}

The second main ingredient for the construction of SSMCs is the cell-free system that constitutes the function-generating molecular machinery. In addition to the reconstruction of DNA duplication and transcription, the most relevant function reconstituted in SSMCs is the synthesis of proteins \cite{28}. By encapsulating the transcription/translation machinery, it is indeed possible to produce a well-folded and active protein inside liposomes. The protein can be an enzyme, a pore-forming membrane protein, a receptor, or a reporter protein. Water-soluble proteins are easily synthesized inside liposomes, whereas the production of membrane proteins is much more troublesome, despite some encouraging reports \cite{9}. This suggests that for the establishment of communication protocols, all those signals requiring a membrane receptor and/or a membrane export device should be for the moment discarded in favor of water-soluble elements.
As cell-free transcription/translation system, two possible kits can be used: \textit{(i)} cell extracts, \textit{(ii)} reconstituted systems. Cell extracts from the bacterium \textit{E. coli} guarantee high protein synthesis yields, and the full compatibility with bacterial proteins that are required to be synthesized by SSMCs, in order to produce a chemical signal or to decode it. However, they suffer of poor characterization. The PURE system, instead, is a fully characterized transcription/translation system reconstituted from purified \textit{E. coli} proteins \cite{20}. Although its production rate is about one third with respect to raw cell extracts, it perfectly fulfills the SB requirements of fully characterized ``standard biological parts{"} (see http://partsregistry.org). Both kits base the transcription step (RNA production) on the T7 RNA polymerase.

\begin{figure}[htbp]
\begin{center}
\includegraphics[width=13cm]{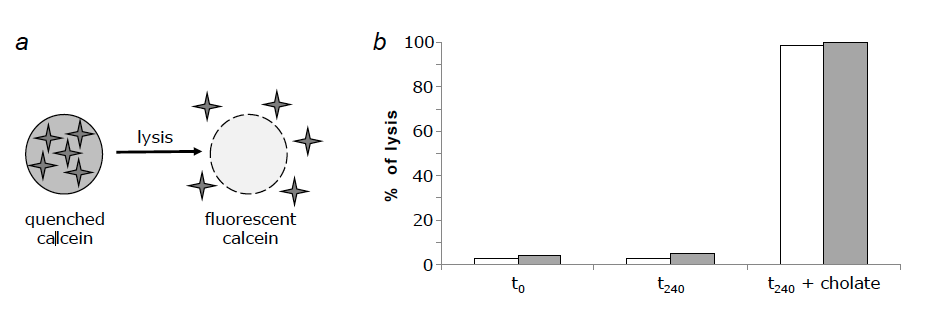}

\caption{(a) Schematic representation of the calcein test used to assess liposome stability. When at high concentration (inside liposomes) calcein is quenched; when at low concentration (released in the medium as a consequence of liposome lysis) calcein is fluorescent. (b) Graph reporting fluorescent emission from calcein-loaded liposomes generated in the bacterial growth medium LB, and incubated for 240 minutes in LB (white bars) or in a bacterial culture grown in LB (grey bars). As a control, cholate was added to induce liposome lysis after 240 minutes incubation ($t_{240}$ + cholate). Maximal fluorescent emission measured is considered as 100\% lysis.}
\label{figure3}
\end{center}
\end{figure}

In preliminary experiments, we have proved that the droplet transfer method can successfully produce protein-synthesizing GVs. At this aim, we have used the green fluorescent protein (GFP) due to its easy detection by fluorescence microscopy. Figure 1d shows that GFP is correctly produced inside several GVs.
Since in our hands the basic translation mechanism works inside GVs prepared by the droplet transfer method, we will next test whether more complex genetic circuits can be reconstructed. Published results clearly show that this is a feasible objective \cite{7,8,21}. The general architecture of minimal signal-synthesizing and signal-detection modules that we will insert into SSMCs are shown in Figure 2, revealing an achievable level of complexity according to the current state-of-the-art.
To confer to the SSMCs the ability of synthesizing a signal molecule able to trigger a response in a natural cell, a DNA vector for the constitutive expression of an AHL synthase will be inserted, together with the PURE system and the AHLÕs substrate, into GVs. In this case the plasmid should contain a transcriptional fusion between a promoter controlled by T7 RNA polymerase and a gene coding for an AHL synthase (Figure 2a). On the other hand, to obtain an SSMC able to respond to the signal sent by the natural cell, the SSMC should be provided with a DNA vector for the constitutive expression of an AHL receptor, and or the receptor-mediated activation of a reporter gene, allowing easy detection of the communication process if taking place. To this aim, the DNA vector should contain a gene coding of an AHL receptor, whose transcription is driven by T7 RNA polymerase, and a reporter gene (\textit{e.g.}, coding for enzymes, fluorescent proteins, light-producing proteins) under the control of a promoter activated by the receptor/AHL complex. In this case, the \textit{E. coli} RNA should be provided, together with the PURE system, to obtain transcription of the reporter gene (Figure 2b).
Another important issue is the physical stability of GVs in the bacterial medium, as well as their possible destabilization due to chemicals or enzymes released by bacteria. Ongoing experiments carried out with conventional liposomes (Figure 3), however, seem to exclude these risks.

\subsection{Bioengineered bacteria}
As previously discussed for the SSMCs, also the natural counterpart of the communication system needs to fulfill specific requirements, depending on the directionality of the communication process envisaged. Clearly, this is needed in this pioneering phase of study, in order to provide a rigorous proof that the synthetic system we are going to experiment behaves as it is expected, being all possible interference eliminated (in real-world cases, the design and the success of synthetic/natural communication will follow the progress of the field). Natural cells (in this case, bacteria) innately fulfill several requirements for establishing natural-to-synthetic and synthetic-to-natural communication channels, while other features need to be generated by bioengineering.
\textit{(i)} Synthesis of a freely diffusible signal molecule stable in the extracellular milieu and able to generate a response in the synthetic cell. Numerous bacteria would be functional in sending a chemical message to a synthetic cell, because bacteria naturally coordinate their social activities at the population level by chemical signaling based on stable signal molecules that can freely diffuse across membranes. In this context, bacteria producing AHLs as signal molecules are suitable candidates for our research program.
\textit{(ii)} Expression of a sensor protein able to perceive the signal molecule sent by the synthetic cell. This is necessary for using natural cell as receivers. AHLs producing bacteria satisfy also this requirement, since they are naturally endowed with receptor systems dedicated to the perception of their own AHLs. Therefore, any bacteria in which intercellular communication is based on a certain AHL, is in principle able to sense the corresponding molecule when synthesized by a neighboring synthetic cell.
\textit{(iii)} Signal decoding capability, \textit{i.e.}, upon stimulus perception, transducing the information contained in the signal molecule to a response element, whose expression/activity must be easily detectable and quantifiable also when at low level. AHL signals are transduced by bacterial cells \textit{via} the AHL receptor itself, which shifts from the inactive to the active state upon signal binding. In the active state, the receptor binds to target promoter regions on DNA, and promotes the transcription of the downstream genes. Therefore, an AHL-producing bacterium should be engineered by inserting in its genome a transcriptional fusion between a promoter activated by the receptor-signal complex and a reporter gene, whose expression levels is easily detectable and measurable. This requirement is technically feasible thanks to the well-known structure of receptors, promoter regions and the good availability of reporter genes (\textit{e.g.}, those coding for enzymes, fluorescent proteins, light-producing proteins).
\textit{(iv)} Be ``signal negative{"}. In the synthetic-to-natural communication design, the synthetic cell should be the only source of chemical signals, and therefore the natural partner (\textit{i.e.}, bacteria) must be deficient/impaired in the synthesis of the signal molecule produced by the synthetic cell to avoid auto-activation of the response. At this aim, the bacteria must be mutagenized (\textit{i.e.}, inactivated) in the gene coding for the enzyme required for the synthesis of the signal molecule, while retaining the ability to respond to it. Also this task is accessible with standard molecular genetic tools.
Finally, as it has been already remarked, natural and synthetic cells must share the same aqueous environment, therefore, they should display reciprocal physical stability. If needed, natural cells should be metabolically active also in synthetic media fitting chemical and physical requirements for synthetic cells formation, stability and functionality. Bacterial metabolic flexibility, demonstrated by studies on species able to communicate \textit{via} AHLs grown in a multitude of mineral media, should permit the identification of adequate chemical and physical parameters allowing co-existence of bacteria and SSMCs. In addition, bacteria should not jeopardize SSMCs physical stability by misinterpreting them as feedstuff. Many bacterial species can use phospholipids (the main constituents of SSMC membranes) as carbon source, by producing enzymes that degrade these macromolecules \cite{26}. Moreover, bacterial species synthesize surfactants that could destabilize liposomal membranes, as well as enzymes with AHLs-degrading activity \cite{1,13}. As previously stated, preliminary data suggest that these undesired events are unlikely in our experimental settings (Figure 3).
At present we are experimentally validating the feasibility of generating engineered bacterial strains endowed with the characteristics previously described. We firstly focus on the use of bacteria as receiver and SSMCs as sender.
Briefly, starting from a bacterium that in nature uses a communication system based on an AHL signal molecule S, we generated a derived mutant strain impaired in the synthesis of this molecule. This strain, named $\Delta S$, is deficient in the synthesis of the molecule S, while normally expressing the cognate receptor RS. Subsequently, we introduced in the $\Delta S$ strain a genetic cassette PS::\textit{lux}, in which the promoter PS, known to be activated by the receptor RS, controls the expression of the \textit{luxCDABE} operon (Figure 4). This operon contains the genes coding for the enzyme luciferase and for the enzymes required for the synthesis of the luciferaseÕs substrate. Therefore, when grown in the presence of exogenous molecule S, the RS receptor is active as a RS-S complex, and drives the expression of the \textit{luxCDABE} operon from the PS promoter, ultimately resulting in light emission. In this way, the response of the bacterium to the molecule S, eventually produced by a SSMC, can be easily detected and quantified by an automated luminometer, also at the microvolumetric scale. We verified this phenomenon by measuring light emission from the bacterial strain  $\Delta S$ PS::\textit{lux} grown in the absence and in the presence of the synthetic molecule S (Figure 5).

\begin{figure}[htbp]
\begin{center}
\includegraphics[width=10cm]{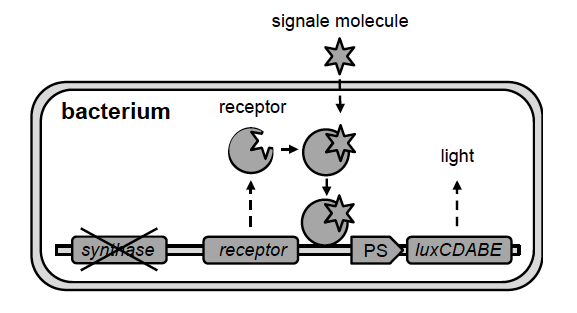}

\caption{Schematic representations of an engineered bacterium unable to synthesize the signal molecule as a consequence of a mutation introduced in the synthase gene, but able to respond to exogenous signal molecule. PS, promoter activated by the signal molecule/receptor complex; \textit{luxCDABE}, operon coding for the enzymes required for light emission.}
\label{figure4}
\end{center}
\end{figure}

Despite being a sensitive and reliable reporter system, luminescence has a main drawback, since it cannot be used to detect activation of the investigated promoter at the single cell level. This is a drawback because it is expected that the amount of signal produced by synthetic cells will be probably low, and localized in the nearby of a single synthetic cell, see also Figure 1c. A refined single cell expression analysis based on the expression of a fluorescent protein could improve the sensitivity of the reporter strain. To this aim, we generated an alternative reporter system in which the \textit{luxCDABE} genes were replaced by the gene coding for the fluorescent protein mCherry (PS::mCherry fusion). The PS::mCherry genetic cassette is expected to allow monitoring the response of the bacterial strain to exogenous molecule S as a function of fluorescence emission.
Preliminary experiments reveal that both the \textit{lux-} and mCherry-based reporter systems were not active when incubated with empty liposomes, while they emitted detectable light or fluorescence signals, respectively, when incubated with liposomes loaded with the molecule S, also demonstrating, by the way, that the signal molecule S can diffuse across the liposome membrane and be perceived by our reporter strain.

\begin{figure}[htbp]
\begin{center}
\includegraphics[width=13cm]{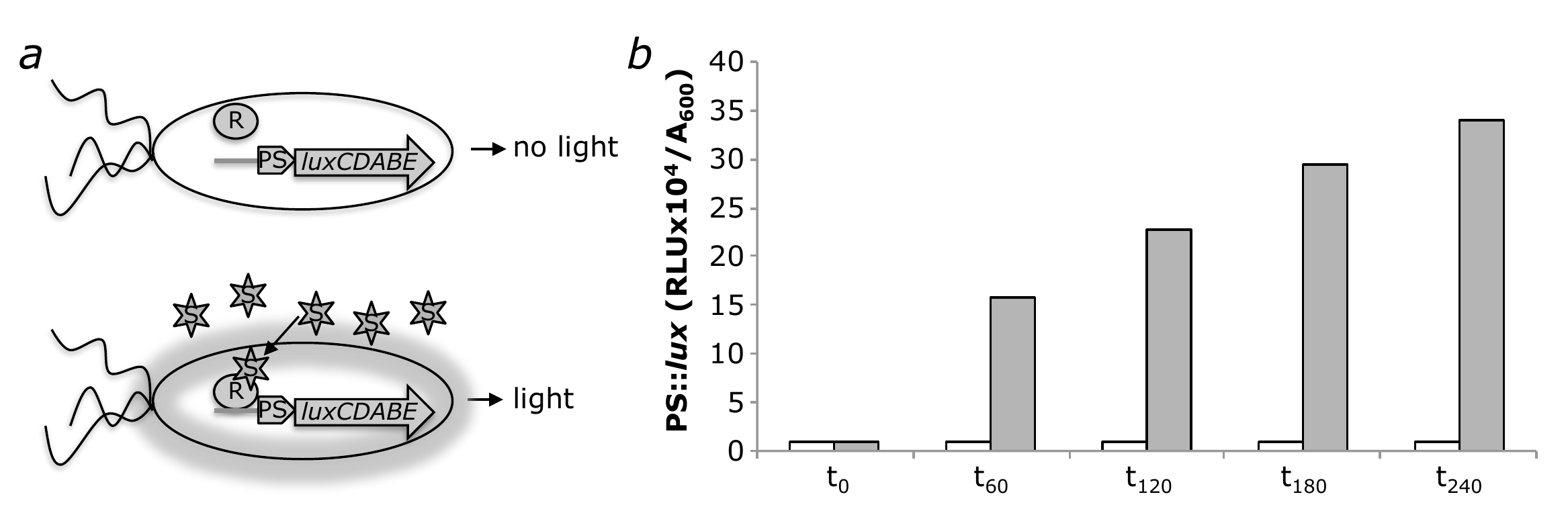}

\caption{(a) Schematic representations of the reporter bacterial system $\Delta S$ PS::\textit{lux}. Light emission is induced by the signal molecule S. (b) PS::\textit{lux} activity measured for 240 minutes in LB (white bars) or in LB supplemented with the signal molecule S (grey bars). The activity of the PS promoter is given as light emission (Relative Light Units, RLU) normalized by cell density ($A_{600}$). RLU and $A_{600}$ were measured every 60 minutes (from $t_{0}$ to $t_{240}$).}
\label{figure5}
\end{center}
\end{figure}

\section{Autopoiesis and AI}
We believe that these advancements in SSMC technology are relevant not only for applications in biotechnological fields, but also for further progress in fundamental science.
SSMC technology develops an experimental approach to the study of minimal life originally implemented in the early 90's by Luisi's group on the basis of the theory of autopoiesis. We think that the experimental scenarios focusing on communication that we are proposing allow an experimental exploration of minimal cognition on the basis of the autopoietic approach. The realization of this exploration would enable synthetic biology to contribute to AI research, in particular with regard to the investigation of minimal cognition and communication that in the last years characterizes AI and, in particular, robotics \cite{15, 22, 27}.

\subsection{Autopoiesis, life and cognition}
Autopoiesis was developed between the 70s and the 90s by Maturana and Varela \cite{14} to tackle the two main questions of cognitive biology: ``what is life?{"} and ``what is cognition?{"} They developed their theory on the basis of three main hypotheses: \textit{(i)} life and cognition are both expressions of the distinctive property of biological systems, that is, ``autopoiesis{"}; \textit{(ii)} autopoiesis (self-production) is the characteristic capability of biological systems to produce and maintain their material identity by themselves, by means of an endogenous processes of synthesis and destruction of their own components (metabolism), which they permanently fuel and realize through active interactions with the environment; \textit{(iii)} autopoiesis is a global or organizational property of living systems, since it relies not on their physico-chemical components taken separately, but in the way in which these components are organized within the systems. These hypotheses allowed Maturana and Varela to address the issue of defining life and cognition as the problem of determining what kind of organization supports the biological behavior of self-production.
They proposed a rigorous solution at the level of the ``minimal cell{"}. This solution consists in the notion of ``autopoietic organization{"}, which intends to significantly contribute to both the disciplinary areas to which cognitive biology belongs.
To the \textit{scientific study of life} Maturana and Varela proposed this notion as a ``synthetic{"} definition of biological systems, which characterizes them not through a list of properties, but through the description of a mechanism able to generate them Ð \textit{i.e.}, the mechanism of self-production of minimal autopoietic systems. The notion of autopoietic organization is being developed within some trends in biology, and within the synthetic exploration of life in its hardware, software and wetware forms. With regard to wetware forms, this notion is at the basis of the above-mentioned synthetic study of minimal life undertaken by Luisi's group.
To the \textit{scientific study of cognition} Maturana and Varela proposed their notion as the theoretical grounding of a biologically convincing model of cognitive systems ``an alternative to the classical model of the computer{"}, ideated within the field of engineering and thus weak from a biological point of view. According to Maturana and Varela, even a minimal autopoietic system is capable of (minimal) ``cognitive behavior{"}. Thanks to the features of its organization, it is able of generating internal operational meanings for perceived external variations. These meanings are expressed in terms of dynamical schemes of self-regulation, which externally appear as actions oriented to conservation (\textit{e.g.}, absorbing a molecule of sugar, overcoming an obstacle...). This meaning generation behavior Ð for Maturana and Varela is the basic cognitive behavior, and grounds what the two researchers called structural coupling with the environment: dynamics of reciprocal perturbations and compensation, in which the autopoietic system continuously generates and associates to exogenous variations operational meanings of self-regulation that allows it to keep its process of self-production in an ever-changing environment.
On the basis of the idea of structural coupling, Maturana and Varela formulated a theory of cognitive interaction between autopoietic systems, and a related theory of communication \cite{14}. According to the latter, communication between two or more autopoietic systems is a dynamics of reciprocal perturbations and compensations, during which each system generates and associates internal operational meanings to the exogenous perturbations produced by the other autopoietic systems. The result of autopoietic communication is conceived by Maturana and Varela as ``behavioral coordination{"}: a mutual and recurrent influence that each system exercises on the other system's behavior not directly, but by stimulating endogenous compensations.
\\
The autopoietic theory of cognition, despite it generated multiple controversial debate, significantly contributed to ground the emerging ``embodied cognitive science{"}, and in particular its ``radical{"} form \cite{3}. Moreover, it has being applied widely in the synthetic exploration of cognition, mainly to implement and test hardware and software models of cognitive processes and systems.

\subsection{Autopoiesis, Synthetic Biology and AI}
Concerning wetware models of cognitive processes and systems, the issue of synthetically implementing and experimentally testing the autopoietic theory of cognition has not been significantly addressed yet. We believe that the SSMC approach we are proposing might allow this kind of experimental exploration of minimal cognition and communication. In this sense, we think that the SSMC approach we presented might generate a synthetic biology program in AI based on autopoiesis.

\section{Concluding remarks}
Top-down synthetic biology largely uses the concepts of logical gates, feedback loops, switches, oscillators, counters to refer to synthetic genetic circuits implanted inside cells, with the aim of controlling the cells behavior, and dealing with the open question ``can chemical signals be manipulated as their electric counterparts{"}?
``Computing{"} with biological parts and devices embodied in bottom-up synthetic cells further expands this field \cite{23}. The construction of cell-like systems with the minimal number of components favors high signal-to-noise ratio, thanks to the elimination of the interference from the background circuitry, that is present in natural cells (a concept known as ``orthogonalization{"}, see \cite{4}). Here we have proposed an experimental plan for endowing SSMCs with elementary genetic devices for sending and receiving chemical information. In this way, synthetic cells might interact with natural cells, or even with other synthetic cells. Although the design here described is rather simple and not yet proved, future advancements might lead to the construction of microscopic machines that are actually soft-wet-micro-robots capable of manipulating and computing signals from biological cells, without the need of translation interfaces.
From a more general viewpoint, however, this synthetic biology approach paves the way to the unprecedented possibility of exploring the dynamics of minimal autopoietic cognition, thanks to physical construction of a suitable model. At the same time, this provides an innovative route to AI.

\section{Acknowledgements}
All authors greatly acknowledge Pier Luigi Luisi (University Roma Tre) for inspiring discussions on synthetic cells and autopoiesis. We are also grateful to Yutetsu Kuruma (University of Tokyo) for supplying the PURE system. This work was supported by a grant to GR from the Ministry of University and Research of Italy (Futuro in Ricerca 2010 - RBFR10LHD1), by grants to LL from Italian Cystic Fibrosis Research Foundation (FFC 14/2010 and FFC 13/2011) and by a grant to LD to develop the Epistemology of the Artificial Project (Dote Ricercatori Regione Lombardia), co-funded by Regione Lombardia and the University of Bergamo.

\bibliography{biblio.bib}

\bibliographystyle{eptcs}
\end{document}